\documentclass[12pt]{article}

\usepackage{amsmath,amsfonts,amssymb,color,graphicx,overpic,} 
\usepackage{slashed}
\usepackage{epsfig}
\usepackage[utf8]{inputenc}
\usepackage{amsmath}
\usepackage{ wasysym }
\usepackage{graphicx}
\usepackage[english]{babel}
\usepackage{graphic x}
\usepackage{relsize}
\usepackage{cite}

\begin{document}


\renewcommand{\arraystretch}{2}

\begin{titlepage}
\rightline{\large February 2016}
\vskip 0.8cm
\centerline{\Large \bf
Solving the small-scale structure puzzles with}
\vskip 0.3cm
\centerline{\Large \bf dissipative dark matter}

\vskip 0.8cm
\centerline{\large Robert Foot\footnote{E-mail address:
rfoot@unimelb.edu.au}}

\vskip 0.2cm
\centerline{\it ARC Centre of Excellence for Particle Physics at the
Terascale,}
\centerline{\it School of Physics, University of Melbourne,}
\centerline{\it Victoria 3010 Australia}
\vskip 0.5cm
\centerline{\large Sunny Vagnozzi\footnote{E-mail address: sunny.vagnozzi@fysik.su.se}}

\vskip 0.2cm
\centerline{\it The Oskar Klein Centre for Cosmoparticle Physics,}
\centerline{\it Department of Physics, Stockholm University, AlbaNova University Center,}
\centerline{\it Roslagstullbacken 21A, SE-106 91 Stockholm, Sweden}

\vskip 0.3cm
\centerline{\it NORDITA (Nordic Institute for Theoretical Physics),} 
\centerline{\it KTH Royal Institute of Technology and Stockholm University,}
\centerline{\it Roslagstullbacken 23, SE-106 91 Stockholm, Sweden}

\vskip 0.8cm
\noindent

Small-scale structure is studied in the context of dissipative dark matter, arising for instance in models with a 
hidden unbroken Abelian sector, so that dark matter couples to a massless dark photon. The 
dark sector interacts with ordinary matter via gravity and photon-dark photon
kinetic mixing. Mirror dark matter is a  theoretically constrained special case where all parameters are fixed except for the kinetic mixing strength,
$\epsilon$. In these models, the dark matter halo around spiral and irregular galaxies takes the form of a dissipative 
plasma which evolves in response to various heating and cooling processes.
It has been argued previously that such dynamics can account for the inferred cored density profiles of galaxies and other related structural features. 
Here we focus on the apparent deficit of nearby small galaxies (``missing satellite problem"), which these dissipative models have
the potential to address through small-scale power suppression by acoustic and diffusion damping. Using a variant 
of the extended Press-Schechter formalism, we evaluate the halo mass function for the special case of mirror dark matter. 
Considering a simplified model where $M_{\text{baryons}} \propto M_{\text{halo}}$, we relate the halo mass function to more directly observable 
quantities, and find that for $\epsilon \approx 2 \times 10^{-10}$ such a simplified description is compatible with the measured 
galaxy luminosity and velocity functions. 
On scales $M_{\text{halo}} \lesssim 10^8 \ M_\odot$, diffusion damping exponentially suppresses the halo mass function, 
suggesting a nonprimordial origin for 
dwarf spheroidal satellite galaxies, which we speculate were formed via a top-down fragmentation process as the result 
of nonlinear dissipative collapse of larger density perturbations. This could explain the planar 
orientation of satellite galaxies around Andromeda and the Milky Way.

\end{titlepage}
 
\newpage

\section{Introduction}
\label{introduction}

The $\Lambda$CDM model has been very successful in describing the large-scale structure of the Universe on
cluster scales and above, and the Cosmic Microwave Background (CMB), see e.g. \cite{planck,kei,wilkinson,tegmark,2dfgrs,blake,sdss,anderson,boss}. Structure
forms hierarchically (bottom-up) 
with the smaller structures forming first 
eventually merging in ever-growing larger structures. In this picture, dark matter (DM) is presumed to be composed of massive collisionless particles
influenced only by gravity with no other (astrophysically significant) forces.
On smaller scales, though, there are indications that something might 
be lacking in this description. For instance, the density profile of galaxies (in particular dwarf galaxies) is known to deviate from the 
cuspy profile predicted from collisionless cold dark matter (CDM hereafter) only simulations, an issue which is known as the ``core-cusp problem" 
(see e.g. \cite{deblok} for a recent review). In addition, the number of nearby small galaxies is found 
to be much lower than expected if DM is both cold and collisionless. This problem is most 
severe for small satellite galaxies (``missing satellite problem'') \cite{klypin,moore}, but also exists 
for field galaxies, see e.g. \cite{Zwaan,kly2,pap}. 
The issue is actually more acute than simply being a local problem, as the observed luminosity and HI mass 
functions exhibit faint-end slopes which are much shallower than those predicted 
by collisionless CDM \cite{zavala}. See also \cite{b3,weinberg} for further discussions on related issues. 

If one insists that dark matter is cold and collisionless then baryonic physics must ultimately resolve the small scale issues, including those
identified above. This is a popular approach adopted in many studies, e.g. 
\cite{vogelsberger,schaye}.  
Here we pursue a very different path. Namely that the small scale
features are taken as evidence that  
dark matter has nontrivial self-interactions cf. \cite{moorex,spergel,m0}.
Here we focus on a
particular scenario where the DM emerges from a hidden sector with an unbroken Abelian $U(1)'$ 
gauge interaction. The associated massless gauge boson, called the dark photon, mediates self  interactions which are also dissipative. 
Models of this kind have been discussed for quite some time in the literature, with mirror dark matter, where the hidden 
sector is exactly isomorphic to the Standard Model (SM), representing the theoretically most constrained 
example \cite{blin,H1,flv,ber1,ignatiev,ber2,m0,sph,sil3,ciar9,cmb2,cmb3}. For a review and more detailed bibliography see e.g. \cite{m1}. 
More generic such models have also been explored in recent years, see 
e.g.  \cite{rich1,berezhiani1,and2,rich2,rich4,rich2x,and1,rich1x,rich5,cyr2,rich6,cyn,fischler1,rich7,sigurdson4,rich8,
petraki2,raidal,reece} for a partial bibliography. The simplest generic model consists of two massive particles, 
usually taken to be fermions, charged under this $U(1)'$ gauge symmetry. The dark sector 
interacts with the SM sector by gravity and via the photon-dark photon kinetic mixing interaction.

Such dissipative DM models can reproduce the successful large-scale structure and CMB predictions of collisionless CDM
\cite{ber1,ignatiev,ber2,cmb2,cyr2,cyn}. 
The dynamics of these models on smaller scales is quite nontrivial. 
In the case of spiral and irregular galaxies, 
the physical picture consists of the DM halo taking the form of a dissipative plasma which evolves in response to various heating and cooling processes. 
At the present time, the halo is presumed to have (typically) reached a 
steady-state configuration where the energy lost due to dissipative interactions is replaced by energy produced by ordinary supernovae;
the precise mechanism involves kinetic mixing-induced processes transferring core-collapse energy into dark photons. 
In fact, it has been argued that the nontrivial dissipative dynamics of such models can 
provide a dynamical explanation for the inferred cored density profile in galaxies, the Tully-Fisher relation, 
and some related structural issues \cite{footexploredb,footexploredc,rich8,footexploredd,footexplorede}. 
However, there has been 
significantly less work addressing other small-scale problems,  including the apparent 
deficit of small galaxies. It has been known for some time that dissipative DM features suppressed power on small scales 
due to dark acoustic oscillations and dark photon diffusion, and therefore has the potential to address this important problem.

Structure formation with suppressed small-scale power has recently been studied in \cite{benson,s13,schneider}. 
It was shown that the extended Press-Schechter (EPS) approach \cite{ps,Bond}, implemented with a sharp-$k$ filter, leads to 
good agreement with simulations. In this paper we apply this formalism to the specific case of dissipative DM to estimate the halo mass function. 
This gives us the number density per unit mass of collapsed objects (dark halos) in the Universe. 
To make contact with observations, we still need to relate the halo mass to some directly measurable quantity, such as luminosity or 
rotational velocity. It would be helpful if we knew the ratio of baryons to dark matter in a given galaxy, but even 
that can be difficult to determine. Although cosmology gives 
us the baryonic mass fraction in the Universe (that is, $\Omega_{b}/\Omega_{m} \simeq 0.15$), on small scales, i.e.  within galaxies, 
baryonic effects such as photoionization and/or supernova feedback (e.g. \cite{bul9}) can potentially cause departures from this 
cosmic value. The size of these baryonic effects are, of course, quite uncertain and 
indications that these effects may not be so significant persist 
(see e.g. \cite{zavala} and references therein). Thus, as a simplified model we consider
$M_b \propto M_{\text{halo}}$ with the proportionality factor tentatively set to $\Omega_{b}/\Omega_{m} \approx 0.15 $.
While somewhat smaller values are preferred by various constraints such as weak lensing and satellite kinematics (e.g. \cite{dutton,mand}),
variation of the $M_b/M_{\text{halo}}$ ratio by a factor of 2 or 3
does not affect our conclusions.

This paper is structured as follows. In Section \ref{dissipative}, we briefly review some aspects of the hidden 
sector $U(1)'$ framework, defining the prototype dissipative DM model. In Section \ref{power}, we consider the matter power spectrum, compute it for some
illustrative examples, 
and briefly discuss the relevant physical processes affecting the growth of structure: acoustic damping 
and diffusion damping. In Section \ref{halo} we give our results for the halo mass function which 
we compare with observations using our simplified $M_b \propto M_{\text{halo}}$ assumption. We also make a 
few speculations regarding the origin of the dwarf spheroidal satellites in the Milky Way and Andromeda systems. 
Our conclusions are given in Section \ref{conclusion}.

\section{Dissipative hidden sector dark matter}
\label{dissipative}

The prototype model for dissipative dark matter presumes the existence of a hidden sector with an unbroken $U(1)'$ gauge interaction. 
The DM then consists of particles carrying $U(1)'$ charge, with the minimal case having two such particles, a ``dark electron'' 
($e_d$) and a ``dark proton'' ($p_d$). The fundamental interactions are described by the following Lagrangian:
\begin{eqnarray}
{\cal L} = {\cal L}_{SM} \ - \frac{1}{4}F^{'\mu \nu} F _{\mu \nu}^{'} + \bar{e}_{d}(iD _{\mu}\gamma ^{\mu} - m_{e_d}) e_d 
+ \bar{p}_{d}(iD _{\mu}\gamma ^{\mu} - m _{p_d})p_d + {\cal L}_{\text{mix}} 
\end{eqnarray}
where ${\cal L}_{SM}$ denotes the $SU(3)_c \otimes SU(2)_L \otimes U(1)_Y$ gauge invariant Standard Model Lagrangian which describes 
the interactions of the SM particles. Also, $F_{\mu \nu}^{'} = \partial _{\mu} A _{\nu} ^{'} - \partial_{\nu} A _{\mu} ^{'}$ [$B_{\mu \nu}  
= \partial _{\mu} B _{\nu}  - \partial _{\nu} B _{\mu}$] is the field strength tensor associated with the $U(1) ^{'}$ [$U(1)_Y$] 
gauge interaction, $A _{\mu}^{'}$ [$B_{\mu} = \cos\theta_w A_\mu + \sin\theta_w Z_\mu$] being the relevant gauge field. The two dark fermions 
are described by the quantum fields $e_d, \ p_d$  and the covariant derivative is: 
$D_{\mu}  = \partial _{\mu} + ig^{'}Q^{'} A_{\mu}^{'}$ (where $g^{'}$ is the coupling constant relevant to this gauge interaction).
While we assume that the dark electron and dark proton have $U(1)'$ charges opposite in sign, 
we make no assumption about their relative magnitude (that is, the ratio $|Q'(p_d)/Q'(e_d)|$ defines a fundamental parameter of the theory).
The self-interactions of the dark electron can be defined in terms of the dark fine structure constant, $\alpha_d \equiv [g'Q'(e_d)]^2/4\pi$. 
The relic abundance of dark electrons and dark protons is presumed to be set by a particle-antiparticle asymmetry (that is, 
the relic abundance of dark antielectrons and dark antiprotons is negligible).
For further discussions of such asymmetric DM models, including possible mechanisms for generating the required asymmetry,
see e.g. \cite{reviewadm,zurekreview}.

In addition to gravity, the dark sector interacts with the SM particles via the kinetic mixing 
interaction between the dark photon and the hypercharge gauge boson \cite{he,flv}:
\begin{eqnarray}
{\cal L}_{\text{mix}} = \frac{\epsilon'}{2} \ B^{\mu \nu} F'_{\mu \nu}
\ .
\label{kine}
\end{eqnarray}
This renormalizable and gauge-invariant interaction also leads to photon - dark photon kinetic mixing. 
The kinetic mixing interaction imbues the dark electron and dark proton with an ordinary electric charge proportional 
to $\epsilon'$, taken to be: $\epsilon e$ and $Z' \epsilon e$, where $Z' \equiv |Q'(p_d)/Q'(e_d)|$ \cite{holdom}. 
The new physics is then fully described by the fundamental parameters: $m_{e_d}, m_{p_d}, Z', \alpha_d$ and $\epsilon$.

Mirror dark matter can be viewed as a special case of the above formalism. That is, mirror DM corresponds to having 
the dark  sector described by a Lagrangian which exactly mirrors that of the SM. In this situation an exact 
and unbroken $\mathbb{Z}_2$ discrete symmetry can be identified. Mirror dark matter is thereby theoretically constrained, 
with $\alpha_d = \alpha$, $m_{e_d} = m_e$, $m_{p_d} = m_p$. Also, $Z' = 1$ for the lightest states (that theory has, of course, a spectrum of dark nuclei).

Cosmological, astrophysical, and some experimental aspects of this and related models have been explored in 
the literature, 
e.g. \cite{cg,rich1,H1,ber1,ignatiev, ber2, cmb2,rich2,rich4,rich2x,rich1x,rich5,cyr2,rich6,foot9,fischler1,cyn,sigurdson4,rich7,petraki2,m1,vogel1,
rich8,raidal,reece,footdiurnal,diurnal,jackson,trodden}. Regarding the  
Early Universe, the thermodynamics of the SM and DM systems can each be described  with a temperature which 
needs not be the same for each sector. Define $T_\gamma$ [$T_{\gamma_D}$] to be the photon [dark photon] temperature, then Big Bang Nucleosynthesis 
and CMB considerations typically constrain the energy density of any exotic radiation component: $(T_{\gamma_D}/T_\gamma)^3 \ll 1$. 
Such considerations will therefore also lead to constraints on the kinetic mixing parameter, $\epsilon$, since kinetic mixing induced processes 
such as $\bar e e \to \bar e_d e_d$ will transfer entropy to the dark sector. That is, even if we started with an effective 
initial condition, $T_{\gamma_D}/T_\gamma \ll 1$, kinetic mixing induced processes can potentially thermalize the SM and dark 
sectors [evolve $T_{\gamma_D}/T_\gamma \to 1$] and thereby violate the constraints on additional energy 
density. For $m_{e_d} \sim$ MeV, the relevant constraint is:  $\epsilon \lesssim 3 \times 10^{-9}$ \cite{foot9,rich8,vogel1}.

Kinetic mixing processes can only be effective for temperatures, $T \gtrsim {\cal M}$, where ${\cal M } \equiv \max(m_e, m_{e_d})$. 
Below this temperature, these processes cease to be important and the ratio $T_{\gamma_D}/T_{\gamma}$ stabilizes. This 
asymptotic value for $T _{\gamma _{_D}}/T _{\gamma}$ was found to satisfy \cite{rich8}:
\begin{eqnarray}
x \equiv \frac{T _{\gamma _{_D}}}{T _{\gamma}} & \simeq & 0.31 \sqrt{\frac{\epsilon}{10^{-9}}} \left (\frac{m _e}{{\cal M}} \right )^{\frac{1}{4}} \ , 
\label{masterformula}
\end{eqnarray}
for parameters in the range $\epsilon \lesssim {\rm few} \times 10 ^{-9}$ and $0.1$ MeV $\lesssim m _{e_d} \lesssim 100$ MeV.

There is an additional source of constraints arising from Early Universe cosmology. Prior to dark recombination, 
the dark electrons and dark protons are strongly coupled to each other and to the dark photons, behaving as a tightly 
coupled fluid. 
Due to the large restoring force from the dark radiation pressure
this fluid undergoes acoustic oscillations which inhibit the formation of structure on scales smaller 
than the sound horizon at dark recombination. 
Too much suppression of structure on scales which are growing linearly today can 
be excluded from observations, leading to the rough bound (e.g. \cite{rich8,trodden,reviewadm}):
\begin{eqnarray}
\epsilon \lesssim 10 ^{-8} \left( \frac{\alpha_d}{\alpha} \right )^4
\left(\frac{m _{e_d}}{\rm MeV} \right)^2 \left ( \frac{{\cal M}}{m_e} \right)^{\frac{1}{2}} \ .
\label{boundse}
\end{eqnarray}
However, some suppression of small-scale power is desirable. Indeed acoustic oscillations, as well as damping due to 
dark photon diffusion, might be the physical mechanisms responsible for the observed suppression of the number of small galaxies relative to 
expectations from collisionless CDM.

\section{Matter power spectrum}
\label{power}

It has been known for some time that this kind of dissipative DM reproduces collisionless CDM on 
large scales with deviations on small scales. The deviations arise from two effects: acoustic oscillations and dark photon diffusion. 
Both effects operate before and during dark recombination, 
and can be characterized by physical scales which depend on the fundamental parameters.

The effects can be described by two-point correlation functions of the density field fluctuations, such as the 
matter power spectrum. This quantity is straightforward to compute, and we will need it later on in order to estimate the halo 
mass function via the extended Press-Schechter formalism. If we define the mass density as $\rho (\mathbf{x})$, then 
inhomogeneities can be described by: $\delta (\mathbf{x}) \equiv [\rho (\mathbf{x}) - \rho_M]/\rho_M$ where $\rho_M$ is the 
mean mass density in the Universe. The power spectrum is related to the Fourier transform of this fractional overdensity, $\delta (\mathbf{k})$, by:
\begin{eqnarray}
\langle \delta (\mathbf{k}) \delta ^{\star} (\mathbf{k'}) \rangle = (2\pi)^3P(k)\delta ^3 (\mathbf{k}-\mathbf{k'}) \ , 
\end{eqnarray}
where the brackets indicate an ensemble average. In practice, the linear power spectrum is determined by solving the set of 
coupled Boltzmann equations for photons, baryons, neutrinos, DM and dark radiation. Since dissipative DM shares many  
similarities with SM baryons, the appropriate Boltzmann equations are straightforward generalizations of those for photons 
and baryons. For the special case of mirror DM they are given in \cite{cmb2} and they may be easily adapted
to the general case.

For our numerical work, we consider a flat Universe with parameters: $\Omega_b = 0.022 \ h^{-2}$,  $\Omega_{\text{dm}} = 0.12 \ h^{-2}$, $h = 0.70$, 
and assume a scale-invariant Harrison-Zel'dovich-Peebles primordial power spectrum of density fluctuations (that is, the spectral 
index is $n_s = 1$). 
The effects of a small tilt, or small changes to $\Omega_b, \Omega_{\text{dm}}, h$, are not important for  our 
considerations. We also consider mirror DM parameters ($\alpha_d = \alpha$, $m_{e_d} = m_e$, $m_{p_d} = m_p$, 
$Z' = \ 1$) \ to illustrate the 
physics.\footnote{
The mirror DM model has a spectrum of dark ``mirror nuclei'' composed 
predominately of mirror hydrogen ($Z'=1,\ m=m_p$) and mirror helium ($Z'=2,\
m=4m_p$). 
It follows that the chosen mirror DM parameters give results for that
model only if mirror hydrogen dominates over mirror helium, i.e. the mirror helium mass fraction is zero ($Y_{He'} = 0$). 
However, the expectation from mirror Big Bang Nucleosynthesis
is that mirror helium dominates over mirror hydrogen \cite{ber1}, with $Y_{He'} \approx 0.9$ for $\epsilon \sim 10^{-9}$ \cite{fc}.
Nevertheless, it turns out that the matter power spectrum and baryonic mass function 
depend weakly on $Y_{He'}$, and we have checked (for $x=0.15$) that the power spectrum (and mass function) results for $Y_{He'}=0$ are in fact 
a good estimate for $Y_{He'} \approx 0.90$. 
}
With these parameters fixed, our results depend 
only on the kinetic mixing parameter, $\epsilon$ [or equivalently, the ratio 
of the dark sector temperature to the visible sector temperature $x$, given Eq.(\ref{masterformula})]. Figure 1 compares the 
matter power spectrum with $x=0$, cosmologically equivalent to collisionless CDM, 
with

\centerline{\epsfig{file=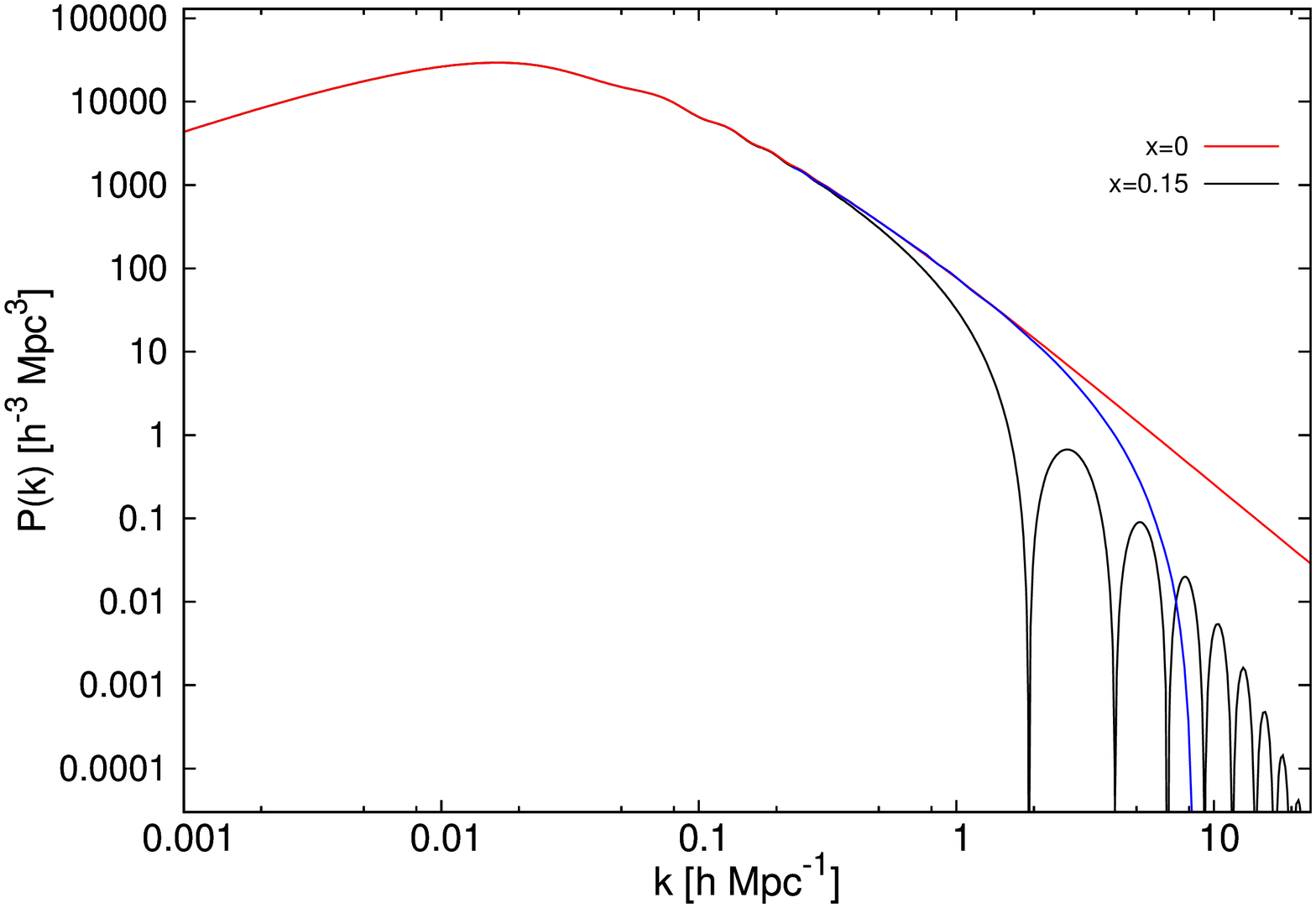,angle=270,width=10.6cm}}
\vskip 0.15cm
\centerline{\epsfig{file=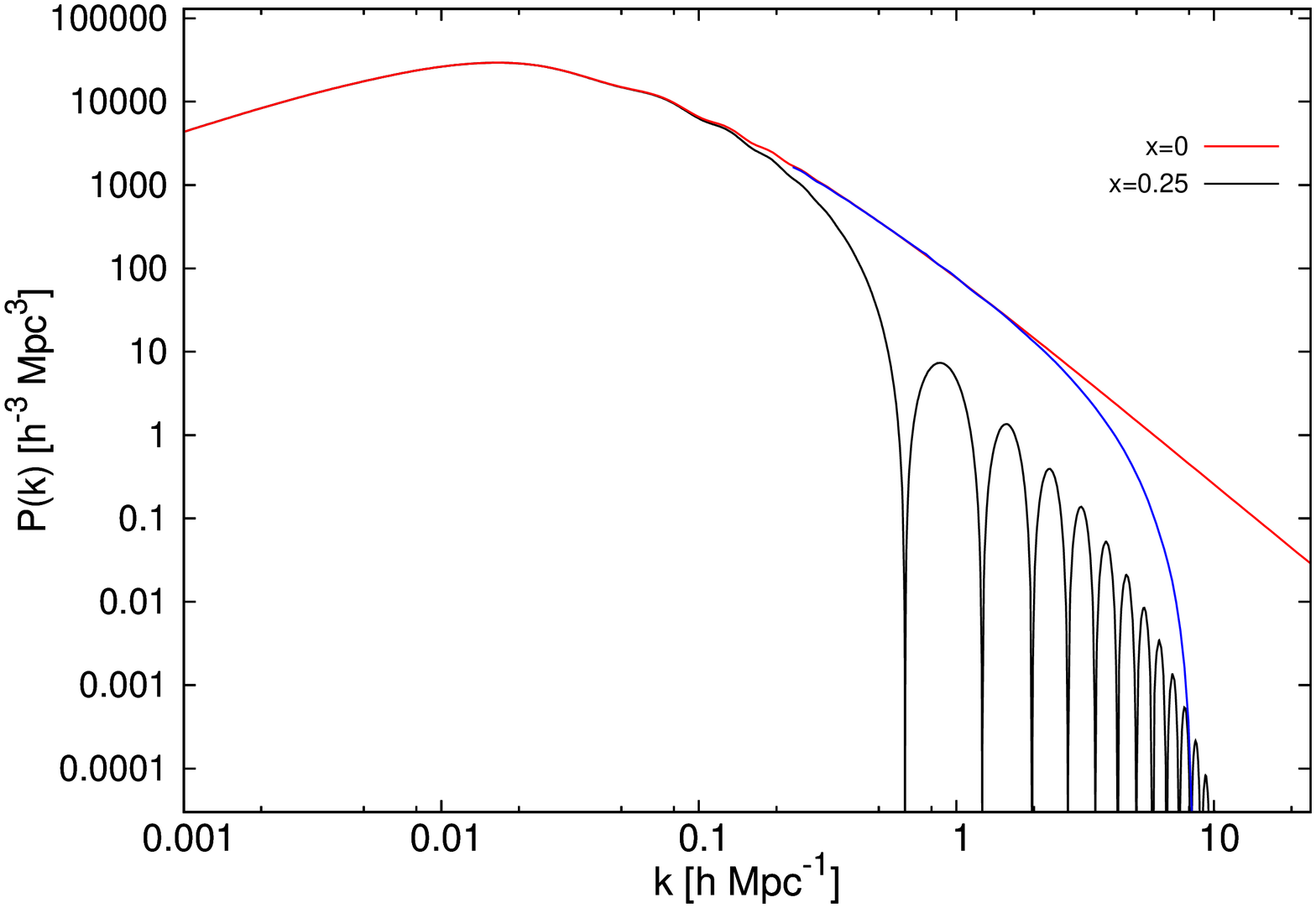,angle=270,width=10.6cm}}
\vskip 0.18cm
\noindent
{\small Figure 1: Matter power spectrum for mirror DM parameters with $x=0$ (top red curves in figures) and 
$x=0.15$ (black curve in Figure 1a), $x=0.25$ (black curve in Figure 1b). These values of $x$ correspond to $\epsilon \simeq 2.3 \times 10^{-10}, \ 6.5 \times
10^{-10}$ respectively, 
from Eq.(\ref{masterformula}). Note that the $x=0$ curve is identical to the standard matter power spectrum for collisionless CDM.
For comparison the power spectrum of a warm DM cosmology, with a thermal equivalent warm DM particle mass of 2 keV, is also shown (blue curves in the two
figures).
}
\vskip 0.95cm

\noindent
our model for two examples. The figure shows 
that the model reproduces the predictions of collisionless CDM on large scales 
with 
departures on small scales.

The modifications to the power spectrum evident in the figures have been previously discussed in the literature \cite{ber1,ignatiev,ber2,cmb2,cyr2,cyn}. 
The coupling of the DM to the thermal bath of dark radiation significantly suppresses the growth of structures on scales below 
the sound horizon at dark recombination. The way it does so can be understood in terms of two important effects: acoustic damping due to 
dark acoustic oscillations (DAOs), and diffusion damping (\textit{collisional damping}) due to dark photon diffusion. Associated to 
each of these effects is a characteristic scale (length scale or, equivalently, comoving wavenumber), which can be expressed in terms of the fundamental
parameters of the model. We provide an analytical estimate of these scales in Appendix A.

Acoustic damping can be understood by recalling that, prior to recombination of the dark electrons and dark protons into 
neutral bound states, the behaviour of the DM can be modelled as that of a tightly coupled fluid. The interplay between gravity and the corresponding 
restoring force determined by dark radiation pressure leads to the phenomenon of DAOs which suppresses the
growth of structure on scales smaller than the size of the sound horizon at the epoch of dark recombination. This effect 
is analogous to the well-known baryonic acoustic oscillations (BAOs), which imprints the visible sector sound horizon 
scale at decoupling ($\sim 105 \ h^{-1} \ \rm Mpc$) on the late-time clustering of matter.\footnote{Dark acoustic oscillations can arise 
more generally in models where the DM couples to relativistic particles. Among these are models where 
the DM couples to neutrinos (see e.g. \cite{shoemaker,escudero}), and photons (see e.g.  \cite{pascoli1,pascoli2}).}

The other effect is due to the tiny but non-negligible dark photon free-streaming length at the epoch of dark recombination. 
Dark photon diffusion acts to erase small-scale inhomogeneities and anisotropies, below the dark photon free-streaming length. 
This effect is known as diffusion damping, and is 
analogous to its visible counterpart known as Silk damping, responsible for the suppression of the high-$l$ acoustic peaks of the 
CMB angular power spectrum \cite{silkapj1968}. Diffusion damping effectively provides an exponentially damping envelope which reduces the power on 
scales below the dark photon mean free path at the dark recombination epoch. This effect is qualitatively similar to the small-scale power suppression in 
warm dark matter (WDM) models. In that case the suppression is due to \textit{collisionless damping}: that 
is, free-streaming of high-velocity DM particles out of initial overdensities, leading to a sharp cut-off 
in power below a critical scale. 

It should be possible to derive constraints on small-scale power suppression from Lyman-$\alpha$ forest considerations (e.g.  \cite{lf1,lf2,seljak}). 
In fact, it has been found that the matter power spectrum for 
$k \lesssim 2 \ h^{-1} \ \rm Mpc$ approximately agrees with that predicted by collisionless CDM (e.g. \cite{seljak}). 
This could be used to infer a rough upper bound:  $k_{\text{DAO}} \gtrsim 2 \ h^{-1} \ \rm Mpc$, which  
for mirror DM parameters [cf. with Eq.(\ref{Ls1})] suggests $x \lesssim 0.15$. This rough bound is subject to a few 
important caveats.  Firstly, there can be significant uncertainties when converting 
the Lyman-$\alpha$ flux power to the linear matter power spectrum \cite{peiris}. Secondly, these analyses typically make 
simplifying assumptions on the shape of the power spectrum (power-law with a running spectral index)  
which could make application of these bounds to models with DAOs problematic \cite{cyr2}.  
Finally, as remarked in \cite{sigurdson4}, to accurately evaluate Lyman-$\alpha$ 
forest constraints on these kinds of self-interacting DM models would ultimately require careful hydrodynamical 
simulations which have not yet been done.

\section{Halo and baryonic mass function}
\label{halo}

Ultimately we would like to describe the structure in the Universe: the distribution of galaxies and their DM halos as a function of their 
masses. To this end, we need to know one-point statistics such as the halo mass function: the number density of collapsed objects per 
unit logarithmic mass in the Universe. 
The extended Press-Schechter (EPS) formalism \cite{ps,Bond} 
is a simple analytic approach which allows us to calculate the halo mass function 
(for a review and more detailed bibliography, see e.g. \cite{psreview}). 
This method assumes linear growth of perturbations and subsequent immediate halo collapse when the mass overdensity reaches a certain 
critical threshold (which is itself derived from an idealised spherical or ellipsoidal collapse toy model).

In the EPS formalism the halo mass function takes the form:
\begin{eqnarray}
\frac{dn}{d\ln M_{\text{halo}}} = \frac{1}{2}\frac{\bar{\rho}}{M}f(\nu)\frac{d\ln \nu}{d\ln M_{\text{halo}}} \, ,
\label{hmf}
\end{eqnarray}
where we have denoted by $n$ the number density of collapsed objects (halos), 
whose mass is given by $M_{\text{halo}}$, $\bar{\rho}$ is the average matter density in the Universe, while the first-crossing distribution 
$f(\nu)$ will be defined shortly. Also, $\nu$ is defined as:
\begin{eqnarray}
\nu \equiv \frac{\delta^2_c}{S(R)} \, ,
\label{peakheight}
\end{eqnarray}
where the critical overdensity for collapse at redshift 0 evaluates to $\delta_c \simeq 1.686$. In Eq.(\ref{peakheight}), 
$S(R)$ is the variance function of the density field smoothed over a length scale $R$
(which corresponds to a mass scale $M$ once a map between the two is given):
\begin{eqnarray}
S(R)  = \frac{1}{2\pi ^2}\int dk \ k^2P(k)|W(k;R)|^2 \, .
\label{variance}
\end{eqnarray}
In Eq.(\ref{variance}), $P(k)$ is the linear matter power spectrum at redshift $z=0$. Also, $W(k;R)$ is the Fourier transform of the filter function, which we
will return to in a moment. Finally, the first-crossing distribution $f(\nu)$ can be obtained by utilizing the excursion-set approach, which follows random
walk trajectories and counts the first up-crossings of the collapse threshold. In the case of ellipsoidal collapse, the first-crossing distribution takes the
form \cite{smt}:
\begin{eqnarray}
f(\nu) = A\sqrt{\frac{2q\nu}{\pi}} \left [ 1 + (q\nu) ^{-p} \right ] e ^{-\frac{q\nu}{2}} \, ,
\label{firstc}
\end{eqnarray}
where $A = 0.3222$, $p = 0.3$ and $q = 1$.

Crucial to the success of the EPS formalism is the choice of filter function $W(k;R)$. For DM models with small-scale power suppression, a sharp-$k$ filter
has been found to reproduce results of simulations \cite{benson,s13,schneider}. The more common choice of real-space top-hat filter instead fails to match
the simulations. 
The reason is readily understood upon inspection of Eq.(\ref{variance}): 
if the real-space top-hat filter is used then 
the variance function becomes insensitive to the power spectrum in the suppressed regime 
(where the power decreases more steeply than $k^{-3}$) 
as the form of the halo mass function is then driven solely by the filter function.

The sharp-$k$ filter is a top-hat function in Fourier space: $W(k;R) \equiv \Theta (1 - kR)$,  where $\Theta$ denotes the Heaviside step function. With this
choice of window function, the form of the halo mass function [Eq.(\ref{hmf})]  reduces to the simple expression:
\begin{eqnarray}
\frac{dn}{d\ln M_{\text{halo}}} = \frac{1}{12\pi ^2}\frac{\bar{\rho}}{M_{\text{halo}}}\nu f(\nu)\frac{P(1/R)}{\delta _c ^2R ^3} \, .
\label{hmfnew}
\end{eqnarray}
The sharp-$k$ filter, unlike its real space counterpart, does not naturally have a mass scale 
$M_{\text{halo}}$ associated to the filter scale $R$. This problem can be resolved by matching 
the expected analytic form for the halo mass function to simulations, which yields: 
\begin{eqnarray}
M_{\text{halo}} = \frac{4\pi \bar{\rho}}{3}(cR) ^3 \, ,
\label{mass}
\end{eqnarray}
where $c \simeq 2.5$ \cite{s13,schneider}. We will use $c = 2.5$ in our numerical work.

\centerline{\epsfig{file=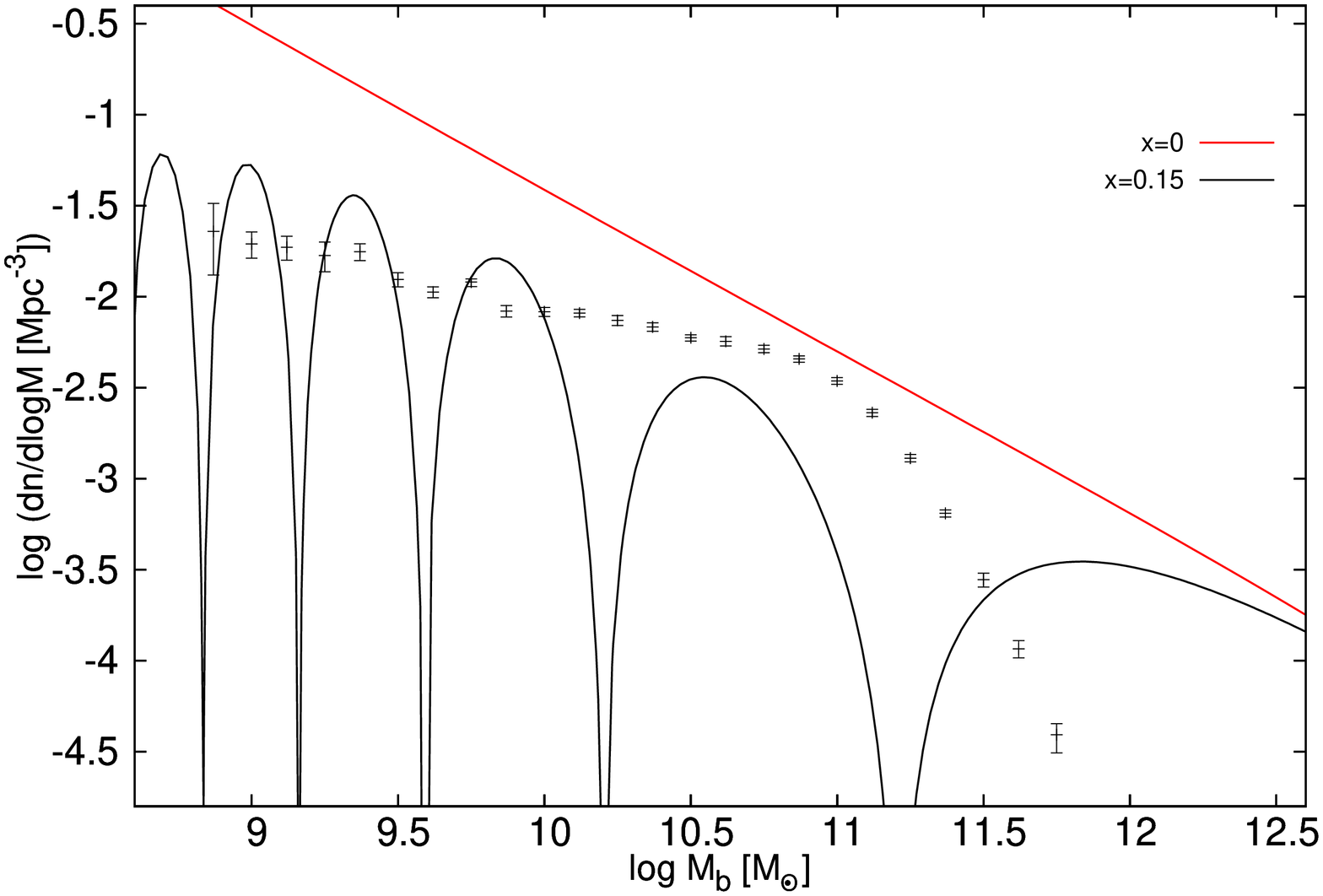,angle=270,width=10.6cm}}
\vskip 0.1cm
\centerline{\epsfig{file=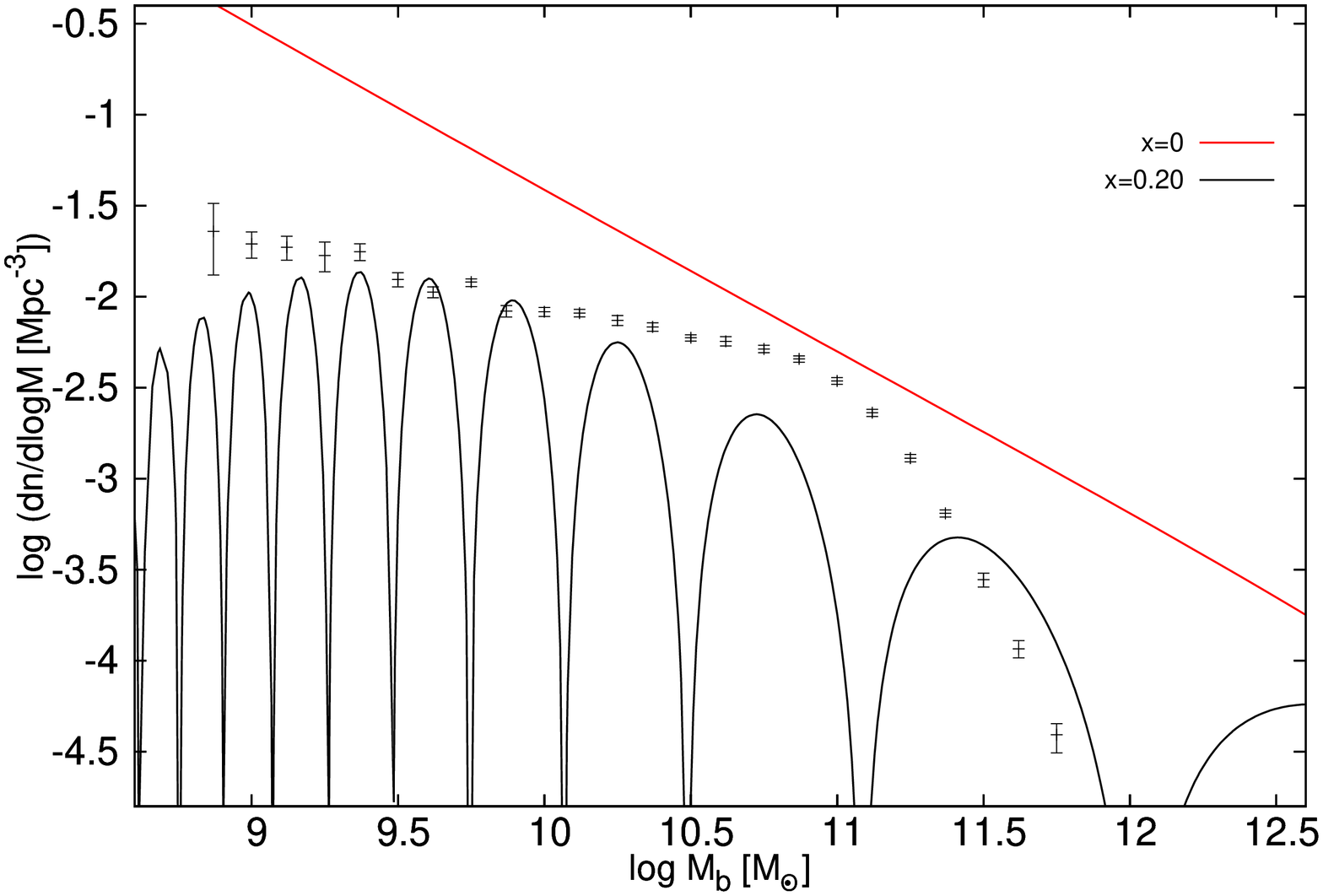,angle=270,width=10.6cm}}
\vskip 0.1cm
\centerline{\epsfig{file=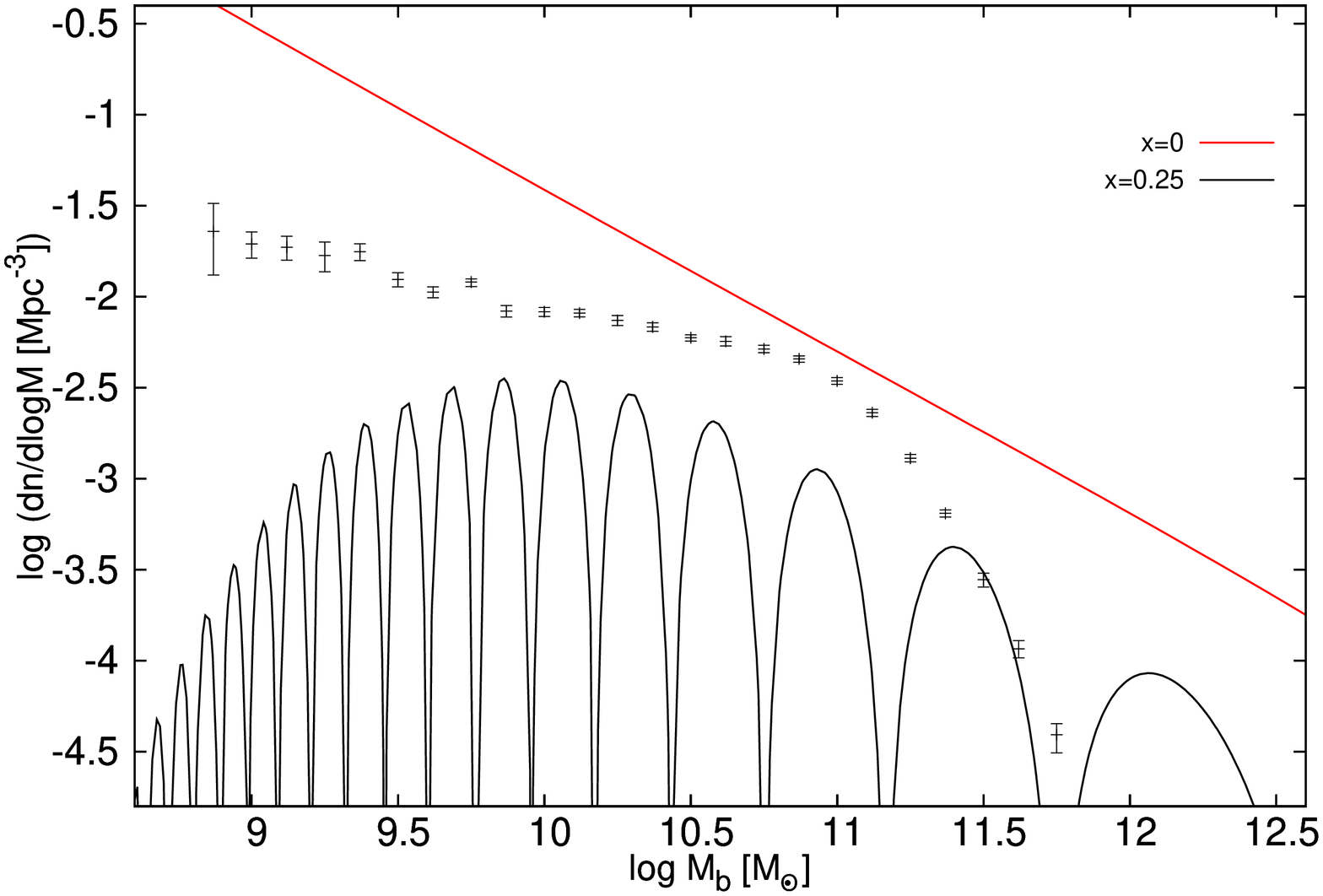,angle=270,width=10.6cm}}
\vskip 0.2cm
\noindent
{\small Figures 2a,2b,2c: Halo mass function, $\log(dn/d \log M_{\text{halo}} \ [\rm Mpc^{-3}])$  versus $\log [M_{b}/M_\odot$] 
with $M_b = 0.15M_{\text{halo}}$ for $x=0$ (top red line, corresponding to $\epsilon = 0$) and $x=0.15,\ 0.20,\ 0.25$ 
(corresponding to $\epsilon/10^{-10} \simeq 2.3, \ 4.2,  \ 6.5$). Data is the galaxy stellar mass function from \cite{bell}. 
} 
\vskip 0.3cm

By using the linear power spectrum computed in the previous section we can use the EPS formalism with sharp-$k$ filter to 
obtain the halo number density in terms of halo mass for the dissipative DM model of Section \ref{dissipative}. We again consider 
mirror DM parameters and the fiducial cosmology as in Section \ref{power}. This serves as a useful 
estimate for mirror DM, as well as providing an illustrative example of the more generic dissipative case.

Given that the halo mass is difficult to measure, it is desirable to relate $M_{\text{halo}}$ to a more directly observable quantity. 
The most readily observable quantities  are connected to the luminous tracers of the DM halos, and this brings us to examine the 
luminosity and rotational velocity functions. We consider a simplified model where the baryon mass content of galaxies, $M_{b}$, 
is related to the halo mass by a simple proportionality relation. That is, $M_{b} \propto M_{\text{halo}}$, and we set the 
proportionality constant to the cosmic value, $\Omega_b/\Omega_m \simeq 0.15$. Such a high value for the baryon to halo mass ratio for 
spiral galaxies appears to be  in some tension (by around a factor of two or three) with weak lensing studies and satellite 
kinematics (e.g. \cite{dutton,mand}).\footnote{However, we have checked that $M_b/M_{\text{halo}} \approx 0.15$ is 
roughly consistent (i.e. typically within no more than a factor of two) with inferences from rotation curves of specific galaxies. 
Taking the NGC 2903 spiral galaxy as an example, $M_{\text{halo}}$ can be estimated from the fit to the rotation curve 
with quasi-isothermal profile \cite{things} (the quasi-isothermal profile is roughly consistent with  expectations assuming dissipative 
halo dynamics \cite{m1,rich8,footexploredd}). This gives a lower bound on $M_{\text{halo}}$ because the radial extent of the 
halo is unknown. Taking the radial extent of the halo to be given by the largest radius for which rotation curve measurements are 
available, i.e. about 30 kpc for NGC 2903, gives the lower limit: $\log M_{\text{halo}}/M_\odot \gtrsim 11.4$. Combining 
this limit with the baryonic mass estimate $\log M_b/M_\odot \simeq 10.5$ \cite{things}, yields a rough limit $M_b/M_{\text{halo}} \lesssim 0.13$. 
Considering the gas rich dwarf DDO 154 as another example, a similar procedure using results from \cite{littlethings} gives the 
upper limit $M_b/M_{\text{halo}} \lesssim 0.10$.}
Naturally, these constraints
on the $M_b/M_{\text{halo}}$ ratio depend on the specific assumptions 
made such as the choice of 
DM halo profile and baryonic mass-to-light ratio, and of course
the total mass of baryons in spiral galaxies is uncertain (see e.g. \cite{italian}). 
In any case variation of the $M_b/M_{\text{halo}}$ ratio by a factor of 2 or 3
does not significantly affect our conclusions.

Figure 2 gives the results for the halo mass function $dn/d\log M_{\text{halo}}$ plotted against $M_b = 0.15M_{\text{halo}}$. 
Notice that within our simplified model where $M_b \propto M_{\text{halo}}$, $d \log M_{\text{halo}} = d\log M_b$, and hence the results 
in Figure 2 are additionally a proxy for the baryonic mass function.
The data  in Figure 2 is the galaxy stellar mass function obtained from \cite{bell} (which also provides a rough estimate for baryons 
given that large galaxies are typically dominated by stars). 

The oscillations apparent in Figure 2 are due to 
dark acoustic oscillations in the matter power spectrum.
These DAOs are not diminished 
at all with the sharp-$k$ filter, but would 
be expected to be diluted to some extent if a more physical filter could be found.
Ultimately, detailed 
simulations should be used to quantify the true magnitude of the DAO effects that might be imprinted on the mass function.\footnote{
Some simulations of models featuring DAOs have been carried out in \cite{sigurdson4}. However, 
the effective halo mass resolution in those simulations
would need to be improved by an order of magnitude or more if one wishes to investigate to full extent the impact of DAOs on the abundance 
of small galaxies.} 
The first few acoustic oscillations at the largest galaxy scales
would present the best chance to detect dark acoustic oscillations.
Actually, our results in Figure 2a and 2b suggest the intriguing possibility that the first dark acoustic oscillation might be connected (or at least
contribute) to the downturn
at ${\rm log}(M_b/M_{\odot}) \approx 11.2$ if $\epsilon \approx 3 \times 10^{-10}$.
On smaller scales though,
DAOs might be difficult to observe as they are expected to be largely smoothed out in the 
process of going from $M_{\text{halo}}$ to $M_b$ to observations. Also, nonlinear 
evolution (which is not accounted for in our linear Boltzmann treatment) would progressively erase DAOs, regenerating power 
on scales which were initially affected by them \cite{sigurdson4}.

\vskip 0.1cm
\centerline{\epsfig{file=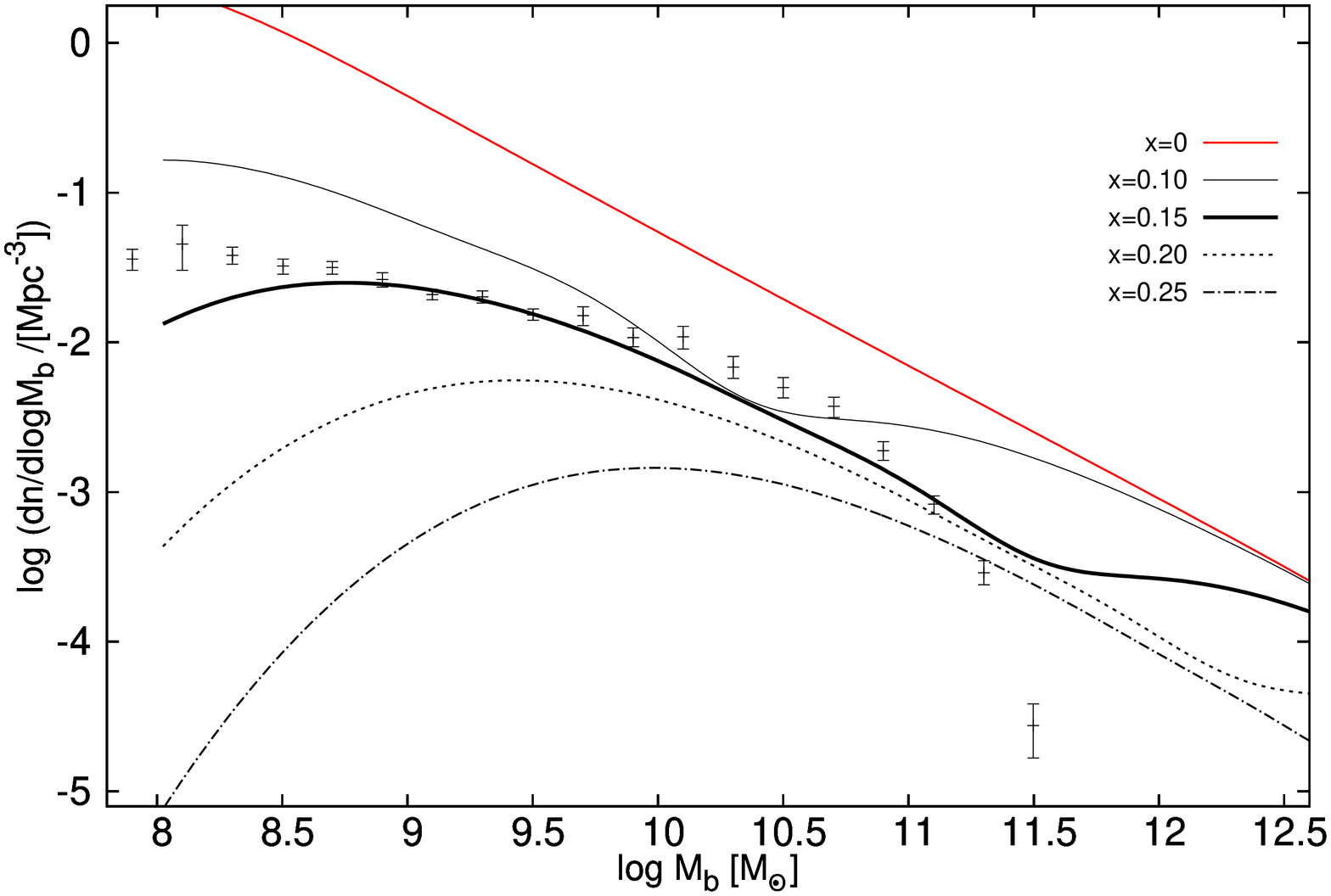,angle=270,width=10.9cm}}
\vskip 0.22cm
\noindent
{\small Figure 3: Baryonic mass function, $\log (dn/d\log M_{b} \ [\rm Mpc^{-3}])$, versus $\log (M_{b}/M_{\odot})$, 
in the simplified model with $M_b = 0.15M_{\text{halo}}$.
Curves top to bottom are for
$x=0$ (red curve) and $x= 0.10,\ 0.15,\ 0.20,\ 0.25$ (corresponding to $\epsilon/10^{-10} = 0, \ 1.04,  \ 2.3 , \ 4.2, \ 6.5$ 
respectively). Data is from \cite{1208}.}
\vskip 1.15cm

For the purposes of the present paper, we shall model these anticipated smoothing effects in a phenomenological way 
by convolving the halo mass function with a Gaussian:
\begin{eqnarray}
\frac{dn}{d \log M_b} \equiv \frac{1}{\sigma \sqrt{2\pi}} \ \int \frac{dn}{d\log M_{\text{halo}}} \ 
e^{-0.5 \left\{ [\log (0.15\times M_{\text{halo}}) - \log (M_b)]/\sigma\right\}^2} \ d \log M_{\text{halo}} \ .
\label{gauss}
\end{eqnarray}
In Figure 3 we plot the baryonic mass function defined in this way, with $\sigma = 0.4$, compared with 
the galaxy baryonic mass function from \cite{1208}. The galaxy baryonic mass 
function includes the baryonic gas component which 
can dominate over stars for the smaller galaxies.
The effect of varying $\sigma$ is shown in Figure 4 for $x=0.15$.
This figure indicates that the effect of variation of the phenomenological parameter $\sigma$ is rather modest,
except perhaps at low $\sigma$ where the oscillation effects start to become noticeable.

An alternative way of comparing the theory to observations is to use the galaxy velocity function rather than 
the baryonic mass function. That is, $dn/d \log V_c$, where $V_c$ is the (circular) rotational velocity of the galaxy. 
This distribution can be estimated from the halo mass function using the baryonic Tully-Fisher relation to relate $M_b$ to $V_c$: 
\begin{eqnarray}
\log [M_b/m_\odot] = s \log [V_c/({\rm km/s})] + \log [A] \ .
\end{eqnarray}
The quantities $s$ and $A$ were found in \cite{lelli} to be $s=3.75 \pm 0.11$ and ${\rm log}[A] = 2.18 \pm 0.23$ from a fit 
to a galaxy sample with accurate distance measurements (we use the central values of these quantities in our numerical work). Given the baryonic Tully-Fisher
relation, the galaxy velocity function is simply:
\begin{eqnarray}
\frac{dn}{d\log V_c} = s \ \frac{dn}{d\log M_b} \ .
\end{eqnarray}
The velocity function defined in this way is shown in Figure 5 (for the same parameter set as per Figure 3).

\centerline{\epsfig{file=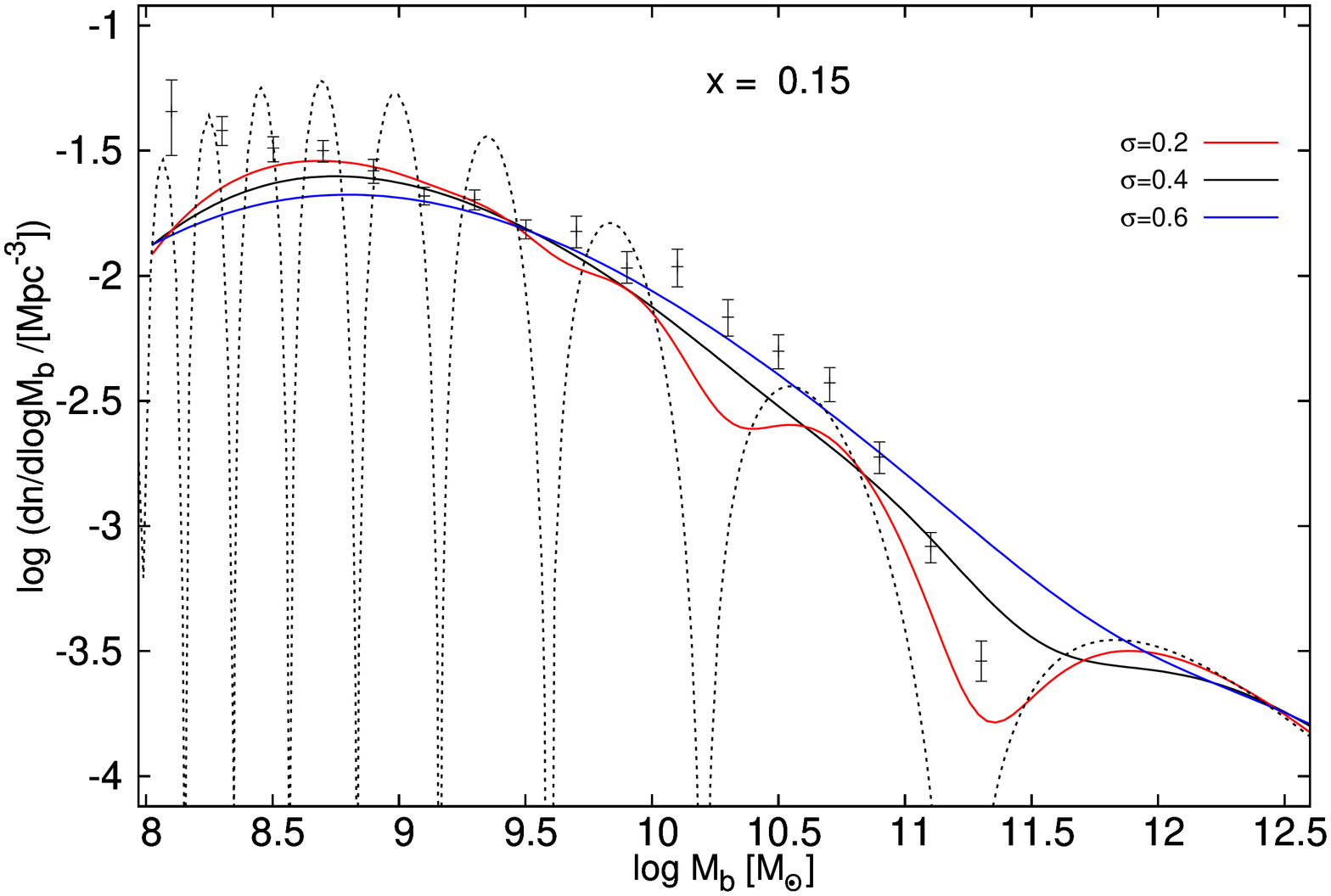,angle=270,width=10.9cm}}
\vskip 0.17cm
\noindent
{\small Figure 4: Baryonic mass function, $\log (dn/d\log M_{b} \ [\rm Mpc^{-3}])$, versus $\log (M_{b}/M_{\odot})$, 
in the simplified model with $M_b = 0.15M_{\text{halo}}$ for $x=0.15$ ($\epsilon  \simeq 2.3 \times 10^{-10}$).
Curves top to bottom are smoothed via Eq.(\ref{gauss}) with
$\sigma = 0.6$ (blue) $\sigma = 0.4$ (black) and $\sigma = 0.2$ (red).  Also shown (dotted line) is the unsmoothed halo mass function.  
Data is from  \cite{1208}.}
\vskip 1.02cm

Also shown in Figure 5 is the measurement of the velocity function using data from the HI Parkes All Sky Survey (HIPASS) \cite{Zwaan}.
Although the theoretical curves give the mass function averaged over the Universe,
the measurements in Figure 5 (as with Figure 3) can only provide the velocity function (mass function) for the smallest galaxies in the nearby region: 
$V \sim $ (5 Mpc)$^3$. Indeed, the effect of cosmic variance on 
the velocity function was examined in \cite{Zwaan}
by subdividing their sample into four quadrants. 
The resulting velocity function for each quadrant exhibited a 
factor of two variation in the normalization, 
the shape, however, was found to deviate significantly less.

In deriving the baryonic mass function in figures 3,4 one should keep in mind that
the results depend significantly on
the simple assumptions adopted, the most important of which is that $M_b \propto M_{\text{halo}}$.  Of course, if there are baryonic physics
processes that lead to gas outflows then this could possibly invalidate this simple assumption.
In addition, dark baryonic physics processes could potentially also occur, especially at early times when
dark star formation is envisaged.
Notice that the baryonic mass function predicted in the $\epsilon = 0$ limit reduces to
that of collisionless CDM, and even in that case it has been argued that inclusion of baryonic physics
could result in a baryonic mass function in good agreement with observations, e.g \cite{vogelsberger,schaye}.
Obviously allowing for such large baryonic (and potentially also dark baryonic) physics effects could therefore
lead to a wide range for the kinetic mixing parameter, $\epsilon$.
Nevertheless, the magnitude of the baryonic physics effects is rather 
uncertain, and might still be relatively minor, e.g. \cite{zavala}. Indeed, direct observational evidence 
for large-scale gas
outflows appears to be absent 
e.g. \cite{martin,eymeren1,eymeren2,lelli1,lelli2,lelli3}, 
although it is possible that large-scale gas outflows occur at early 
times when observations are currently limited.
Even if large outflows occur, there remain serious challenges in reconciling the steep galaxy
stellar mass function with observations of dwarf galaxies, e.g. \cite{ferrero,pen1207}.
At any rate, if the baryonic effects do turn out to be minor, then the baryonic mass function
would be a direct probe the dark sector physics, which is the optimistic case we have considered.

\vskip 0.7cm
\centerline{\epsfig{file=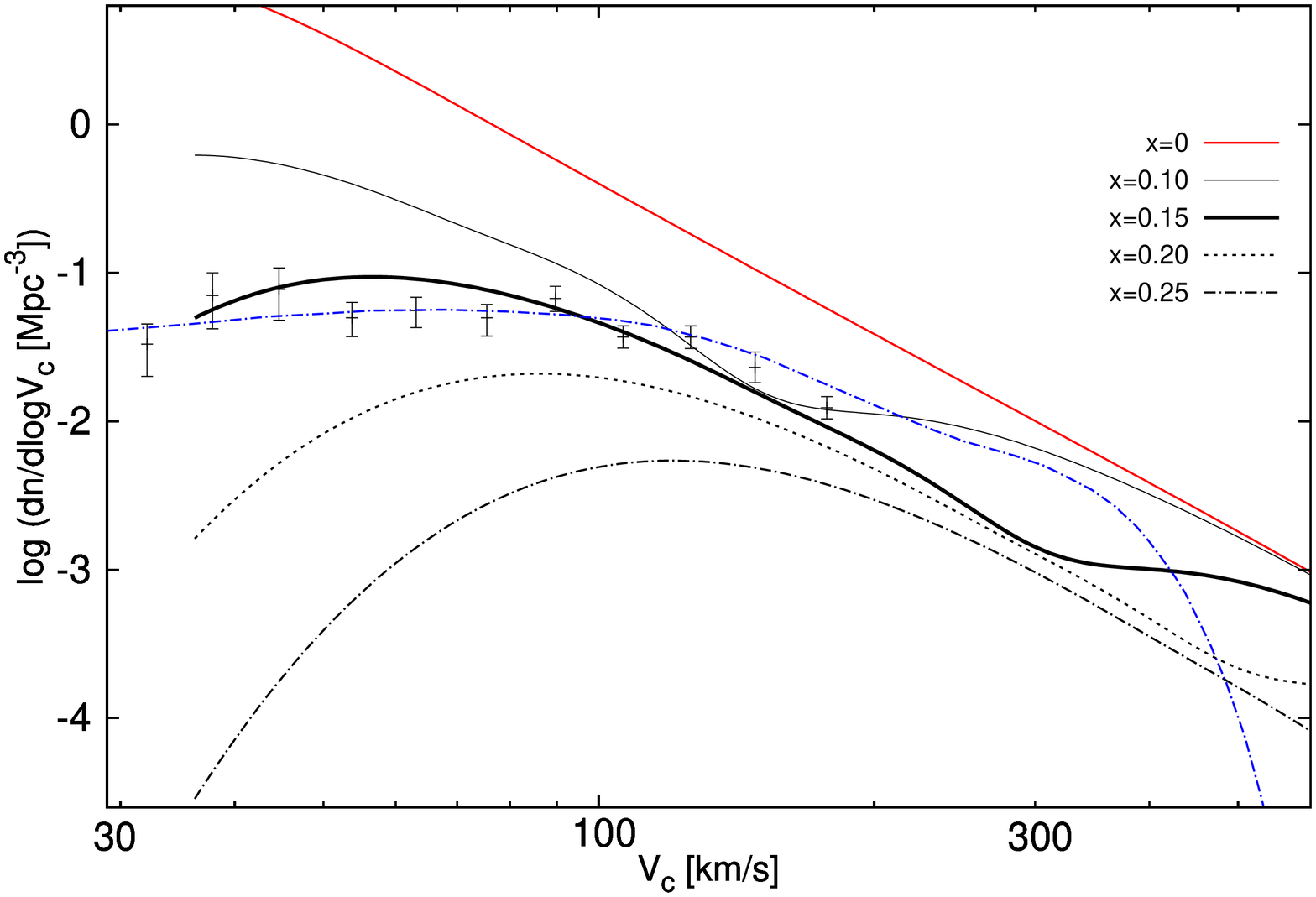,angle=270,width=10.9cm}}
\vskip 0.3cm
\noindent
{\small Figure 5: The velocity function $\log (dn/d\log V_c \ [\rm Mpc^{-3}])$ versus $V_c [\rm km/s]$ 
for dissipative dark matter. 
The top to bottom lines are for $x=0$ (red) and $x= 0.10, \ 0.15, \ 0.20, \ 0.25$ 
(corresponding to $\epsilon/10^{-10} \simeq 0, \ 1.04, \  2.3, \ 4.2, \ 6.5$ respectively).
Data points are the 	HIPASS data for $V_c < 200$ km/s (gas rich sample), while the blue dashed dotted line
is an estimate of the total galaxy count (i.e. including the most massive galaxies) \cite{Zwaan}.}
\vskip 1.4cm

Our results indicate that the considered dissipative DM model can provide a rough explanation of the shape and normalization of the 
measured galaxy luminosity and velocity
functions.
Furthermore, $x \approx 0.15$, that is, $\epsilon \approx 2\times 10^{-10}$
[by Eq.(\ref{masterformula})] is implicated for mirror DM parameters. 
This of course assumes that baryonic and dark baryonic physics do not play major roles
in modifying the assumed $M_b \propto M_{\text{halo}}$ scaling, which is an important caveat as
discussed above.
Allowing for the various uncertainties, suggests a preferred parameter range of roughly $x = 0.10 - 0.20$ or $\epsilon/10^{-10} = [1.0 - 4.2]$. 
Such $\epsilon$ values are consistent with the kinetic mixing strength required by the
dynamical halo model of \cite{sph,footexploredb,rich8}. Recall that, in that picture, 
the halo takes the form of a dissipative plasma which evolves in
response to various heating and cooling processes. At the current epoch DM halos around spiral and irregular galaxies are (typically) 
presumed to have reached a steady-state configuration where heating and cooling rates locally equilibrate. 
The heating is presumed to arise from ordinary core collapse supernovae, with kinetic mixing playing a critical 
role in converting a significant fraction of the supernova's 
core collapse energy into dark photons. In order for this heating to be dynamically important, $\epsilon \gtrsim 10^{-10}$ is required.

The number of large galaxies, i.e. with $M_b \gtrsim 2 \times 10^{11} \ M_\odot$, is observed to be exponentially suppressed. In the case of
collisionless CDM this downturn is believed to be caused by inefficient cooling and energy injection 
from central super-massive black holes (see
e.g. \cite{croton}). It is likely that these effects play a role in the dissipative DM case too. 
However, the results for $\sigma = 0.20, \ x=0.15$
shown in Figure 4
suggest that it is possible that the first dark acoustic 
oscillation might also be connected with the dip at ${\rm log} (M_b/m_{\odot}) \approx 11.2$\ .
More subtle dissipative DM effects could also contribute, e.g. if
the halo cooling timescale were sufficiently long then it could slow the formation of such large halos ($M_{\text{halo}} \gtrsim 10^{12} \ M_\odot$)
and thereby affect the rate of large galaxy formation.\footnote{Another possibility, especially relevant for the mirror DM special case, is that the
ionization state of the 
halo undergoes a transition at $M_{\text{halo}} \sim 10^{12} \ M_\odot$. In the mirror DM model, the 
halo contains mirror ``metal''  components possibly dominated by mirror oxygen. These metal components play a critical role in halo
dynamics as their photoionization provides the mechanism for transferring supernovae core-collapse energy to the halo.
However the mirror oxygen component is estimated to become fully ionized 
at $T_{\text{halo}} \approx 0.5$ keV which corresponds roughly to $M_{\text{halo}} \sim 10^{12} \ M_\odot$ given our 
assumed $M_b = 0.15M_{\text{halo}}$ and the connection between $T_{\text{halo}}$ and $V_{c}$ in this model (see for instance \cite{m1}). 
This could result in a significant reduction in halo heating as photoionization becomes less effective at transferring 
the energy produced by ordinary supernovae to the halo. In the absence of sufficient heating the dissipative 
halo would  contract perhaps triggering AGN activity and/or dark star formation, which might result in a limiting galaxy scale.} 


At very low halo masses, $M_{\text{halo}} \lesssim 10^{8} \ M_\odot$, power is exponentially cutoff due to 
diffusion damping (for mirror DM parameters). The number of very small galaxies is thereby expected to be strongly suppressed. The halo mass 
function gives the number of collapsed  
halo objects to be compared with field galaxies (or clusters on larger scales). Although 
we haven't explicitly calculated the number of satellites around a host (conditional mass function), it is clear that the 
number of satellites will also be strongly suppressed. In fact, the model can potentially ``over solve'' the 
missing satellite problem. However, the small satellites might have a very different history to field galaxies. Rather than evolving out 
of primordial density perturbations, small satellite galaxies might have formed more violently in a top-down process.  
In fact, we will argue below that if dark matter is dissipative then this provides a simple explanation 
for the observed planar and co-rotating distribution of satellite galaxies around
the Milky Way \cite{paw}, Andromeda \cite{ibata} and possibly other hosts \cite{lewis}.

In dissipative DM models, a region with $\delta({\bf x};R) > \delta_c$ can undergo gravitational collapse and cool forming a dark disk. 
Baryons can also collapse, potentially forming a separate disk. Of course there can be some amount of baryons 
in the dark disk (and vice versa). In the model of \cite{sph,footexploredb,rich8}, the dark disk is eventually heated, and any DM not within 
compact dark matter objects (dark stars) can thereby evolve to form the roughly spherical dark halo. Returning to early times though, 
when the more massive dark disk was forming, the complicated nonlinear collapse process is not expected to be uniform and  perturbations 
near the edge of the disk could ultimately break off and provide the seeds of small satellite galaxies. Despite the complexity of this picture, 
it suggests a very simple explanation for the observed planar and co-rotating distribution of dwarf spheroidal galaxies around  
the Milky Way and Andromeda.\footnote{One can further speculate that the observed polar orientation of the satellite planes 
with respect to the baryonic disk is a consequence of the nontrivial dynamics.   
Dark supernovae can be 
a powerful source of SM photons (due to the kinetic mixing interaction) which can heat the baryonic disk at early times providing 
a pressure force. While gravitational attraction between the two disks tend to make them merge (cf. \cite{rich6}), 
the pressure force from the disk heating can work in the opposite manner and can potentially  overwhelm the gravitational force. 
If this is the case, then the baryonic disk would be expected to evolve until it aligns with polar orientation.} 
These satellites broke off from 
the dark disk which formed at very early times, thereby preserving this planar and co-rotating structure.\footnote{A related 
but still distinct possibility discussed in \cite{thindisk,dddm6} is 
that the satellites broke off from the baryonic disk as a result of an ancient merger event \cite{hammer1,hammer2,hammer3}. Galaxies produced in 
this way are known as tidal dwarf galaxies. 
At the present time though, it is unclear whether the dwarf spheroidal satellites of the Milky Way and Andromeda could be interpreted 
as tidal dwarf galaxies \cite{bf,pawlowski}.}  
If they did indeed form in this way then there will be other ramifications.
For example, they may have  
a much larger proportion of DM when compared with bottom-up forming galaxies.

\section{Conclusion}
\label{conclusion}

In this paper, we have studied the abundance and clustering of small-scale structure in the context of dissipative dark matter models. 
Such models feature suppressed power on small scales due to acoustic damping and diffusion damping before and around the epoch of 
dark recombination. We estimated the damping scales analytically and initially studied their effect on the clustering of matter by computing the 
late-time matter power spectrum. Through this small-scale power suppression, these models have the potential to 
address the apparent deficit of nearby small galaxies, including the ``missing satellite problem", as they naturally 
reduce the abundance of structure below the damping scales.

Subsequently, we explored the impact of the power suppression on the abundance of small-scale structure by using a 
variant of the extended Press-Schechter formalism. Within this formalism we estimated the halo mass function for the particular parameter set corresponding 
to mirror dark matter. This serves as a good estimate for mirror DM, as well as providing a  useful  example illustrating possibilities in 
more generic dissipative models. The halo mass function gives the number density of collapsed objects as a function of halo mass. 
Unfortunately, the halo mass itself is difficult to directly measure, so we considered the luminous tracers 
of the dark matter halos. To make contact with such observations, we considered a simplified model where $M_b \propto M_{\text{halo}}$, 
and set the proportionality constant to the cosmic value, $\Omega_b/\Omega_m \simeq 0.15$. This allowed us to connect the host 
halo mass to properties of its tracers such as luminosity or rotational velocity.

With mirror DM parameters set, the luminosity or rotational velocity functions depends only on one parameter: the kinetic 
mixing strength $\epsilon$. Acoustic damping due to DAOs can supply moderate suppression of structure on galactic scales, which 
nevertheless is sufficient to yield a baryonic mass function which compares reasonably well with the observations, 
provided $\epsilon \approx 2 \times 10^{-10}$. This value of $\epsilon$ is consistent with the previously estimated range of $\epsilon$ favoured 
in dissipative halo dynamics: $\epsilon \gtrsim 10^{-10}$ \cite{sph,footexploredb,rich8}. Such a kinetic mixing strength also makes the 
model interesting for direct detection experiments which can probe dissipative DM via electron and nuclear recoils 
(e.g. \cite{jackson} and references there-in). A particularly distinctive signature arises due to the
capture of DM within the 
Earth. The consequent shielding of a detector from part of the halo DM wind results in  a large diurnal modulation effect 
which is enhanced for a 
detector located in the Southern Hemisphere
\cite{footdiurnal,diurnal,jackson}.

The predicted baryonic mass function  falls much more steeply below the diffusion damping scale. For mirror DM parameters, this indicates 
that the abundance of very small galaxies, i.e. with $M_{\text{halo}} \lesssim 10^8 \ M_\odot$, is severely reduced. This in turn 
suggests a very different origin for the dwarf spheroidal satellite galaxies. They may have formed via a top-down process, as 
the result of nonlinear dissipative collapse of larger density perturbations. This origin might also explain some of their 
peculiar features, including the fact that they preferentially orbit in a planar distribution around their host (for the Milky Way and 
Andromeda systems). This plane should align with a remnant dark disk embedded in the dark halo of the host galaxy.

In the present work we computed the baryonic mass function at the present time ($z$=$0$).
Future work could examine the redshift dependence of the observables we considered, which could 
easily be studied within the extended Press-Schechter formalism. Also, we have evaluated the abundance and 
clustering of small-scale structure for mirror dark matter parameters only. It should be straightforward to extend this study to 
more generic dissipative dark matter models, in particular examining the effects of the variation of all the fundamental parameters
associated with the dissipative dark matter model defined in Section \ref{dissipative}.
Understanding the properties of satellite galaxies seems to be a more difficult proposition if these very small galaxies are 
indeed formed from a top-down nonlinear fragmentation process. Presumably, 
a full understanding of this problem would require simulations to be performed. Finally, we emphasise again that 
direct detection experiments will provide the critical test for these models.


\section*{Appendix A - Damping scales}

Here we provide approximate analytical expressions for the acoustic damping (DAO) and diffusion damping (dark Silk) scales. 

\subsection*{A.1 - Acoustic damping scale}

The acoustic damping scale is given by the sound horizon at the epoch of dark recombination: 
\begin{eqnarray}
L_{\text{DAO}} \simeq \int _0 ^{\eta _{\text{dr}}} d\eta \ c_D(\eta) 
= \int _{z_{\text{dr}}} ^{\infty} dz \ \frac{c_D(z)}{H(z)} \ . 
\label{s88}
\end{eqnarray}
Here $\eta$ denotes conformal time [$d\eta \equiv dt/a(t)$], $z$ is the corresponding redshift, $\eta _{\text{dr}}$ is the value of conformal time at dark
recombination and $H(z)$ is the Hubble parameter. Also, $c_D$ denotes the sound speed in the dark sector and is given by:
\begin{eqnarray}
c_D(z) = \frac{1}{\sqrt{3}} \left [ 1 + \Delta (z) \right ] ^{-\frac{1}{2}} \ , 
\end{eqnarray}
where:
\begin{eqnarray}
\Delta (z) &\equiv & \frac{3\rho_{\text{dm}}}{4\rho_{\gamma_{_D}}} =  \frac{B}{1+z}  \ .
\label{delta}
\end{eqnarray}
Here, $B \equiv  45 \Omega_{\text{dm}} \rho_{\text{crit}}/(4\pi^2 x^4 T_0^4)$, $\rho_{\text{crit}} \equiv 3H_0^2/(8\pi G)$ is the critical density, $T_0
\simeq 2.7256$ K is the current CMB temperature, and we have used $T_{\gamma} (z) = T_0 (1 + z)$. Also, the ratio between the temperatures in the two sectors,
$x$, is given in Eq.(\ref{masterformula}). 

The acoustic damping scale, Eq.(\ref{s88}), depends on the redshift of dark recombination, $z_{dr}$, which can be estimated in terms of the fundamental
parameters of our model: 
\begin{eqnarray}
\frac{1}{1+z_{\text{dr}}} = \frac{T_0}{T_{\text{dr}}} = \frac{x T_0}{T_{\text{dr}}^{'}} \simeq \frac{2 x \xi T_0}{{\alpha_d}^2 m_{e_d}} \ .
\label{zdr}
\end{eqnarray}
Here we have related the dark recombination temperature, $T'_{\text{dr}}$, to the $e_d$(s)-$p_d$ bound state binding energy $I' \approx {\alpha_d}^2m_{e_d}/2$
by $T'_{\text{dr}} = I'/\xi$.  The parameter $\xi$ depends only logarithmically on the fundamental parameters and  was estimated to be $\xi \approx 40$ in
~\cite{rich8}.

The integral, Eq.(\ref{s88}) can be analytically solved. Using $H(z) = H_0 (1 + z)^2 \sqrt{\Omega_r}$, appropriate for the radiation dominated era (the
redshifts of interest satisfy $z \gtrsim z_{\text{dr}} \gtrsim z_{\text{eq}}$), we find:
\begin{eqnarray}
L_{\text{DAO}} = \frac{2}{H_0 \sqrt{3\Omega_r} B} \left( \sqrt{1 + y}  - 1 \right) \ ,
\label{lstarnew}
\end{eqnarray}
where:
\begin{eqnarray}
y \equiv \frac{B}{z_{dr}+1} \approx  87 \left( \frac{10^{-9}}{\epsilon}\right)^{3/2} \left( \frac{{\cal M}}{m_e} \right)^{3/4} \left( \frac{\alpha}
{\alpha_d}\right)^2 \left( \frac{m_e}{m_{e_d}}\right) \ .
\end{eqnarray}
Evidently, for a large range of parameter space, $y \gg 1$, and Eq.(\ref{lstarnew}) reduces to:
\begin{eqnarray}
L_{\text{DAO}} &\approx & \frac{2}{H_0 \sqrt{3\Omega_r B (z_{dr}+1)}}  \nonumber \\
& \approx & 8.6 \left( \frac{\epsilon}{10^{-9}}\right)^{5/4} \left( \frac{\alpha}{\alpha_d}\right) \left(\frac{m_e}{{\cal M}}\right)^{5/8}
\left(\frac{m_e}{m_{e_d}}\right)^{1/2} \ h^{-1} \ {\rm Mpc} \ .
\label{Ls1}
\end{eqnarray}
The corresponding associated comoving wavenumber is $k_{\text{DAO}} \approx \pi/L_{\text{DAO}}$, and one can easily check that it is consistent with our
numerical results shown in Figures 1, 2.
%
%

\subsection*{A.2 - Diffusion damping scale}

On scales below the dark photon mean free path, dark diffusion damping is efficient in erasing the DM fluctuations. 
The physics is virtually identical to the well studied case of photon diffusion. Following standard arguments (e.g. \cite{dodelson}),
the diffusion scale, $L_{\text{DSD}}$, can be estimated to be:
\begin{eqnarray}
L_{\text{DSD}} \approx \pi \left [ \int _0 ^{\eta _{\text{dr}}} d\eta ' \ \frac{1}{6(1 + \Delta)n_{e_d}\sigma _{T_d}a(\eta ')} \left ( \frac{\Delta ^2}{1 +
\Delta} + \frac{8}{9} \right ) \right ] ^{\frac{1}{2}} \ ,
\end{eqnarray}
where $\sigma _{T_d} = 8\pi  {\alpha_d}^2/(3m_{e_d}^2)$ is the dark Thomson cross-section. This integral evaluates to:
\begin{eqnarray}
L_{\text{DSD}} \approx \pi \left [ \frac{m_{p_d}}{18H_0\sqrt{\Omega _r}\Omega_{\text{dm}}(1 + z_{\text{dr}}) ^3\rho _{\text{crit}}\sigma _{T_d}} \right ]
^{\frac{1}{2}} \ ,
\end{eqnarray}
where we have made the assumption that $m_{e_d} \ll m_{p_d}$. Using Eqs.(\ref{masterformula},\ref{zdr}), $L_{\text{DSD}}$ can be expressed in terms of the
fundamental parameters:
\begin{eqnarray}
L_{\text{DSD}} & \approx & 0.7 \left( \frac{\epsilon}{10^{-9}}\right)^{3/4} \left( \frac{\alpha}{\alpha_d}\right)^4 \left(\frac{m_e}{{\cal M}}\right)^{3/8}
\left(\frac{m_e}{m_{e_d}}\right)^{1/2} \left (\frac{m_{p_d}}{m_p} \right )^{\frac{1}{2}} \ h^{-1} \ {\rm Mpc} \ .
\label{Ld1}
\end{eqnarray}
For mirror DM parameters, $L_{\text{DSD}} < L_{\text{DAO}}$, however more generally, $L_{\text{DSD}} > L_{\text{DAO}}$ is possible. The associated comoving
wavenumber is $k_{\text{DSD}} \approx \pi/L_{\text{DSD}}$, and like $k_{\text{DAO}}$ is also consistent with our numerical results shown in Figures 1,2.

%
%

The length scales $L_{\text{DAO}}$ and $L_{\text{DSD}}$ and corresponding comoving wavenumber scales are 
derived in linear perturbation theory (i.e. prior to the nonlinear collapse phase). The mass scales that 
correspond to these length scales are
roughly: $M_{\text{DAO}} \sim \pi \rho_{\text{crit}} \Omega_m L_{\text{DAO}}^3/6$,  $M_{\text{DSD}} \sim 
\pi \rho_{\text{crit}}\Omega_m L_{\text{DSD}}^3/6$. In terms of the wavenumber, $k$, the correspondence is
approximately [from Eq.(\ref{mass})]:
\begin{eqnarray}
M(k) \approx 3\times 10^{12} \left ( \frac{k}{\rm Mpc^{-1}} \right ) ^{-3} \ M_\odot \ .
\end{eqnarray}
%


\vskip 1cm

\noindent
{\large \bf Acknowledgments}
\vskip 0.2cm
\noindent
RF is supported by the Australian Research Council and SV is supported by the Vetenskapsr\r{a}det (Swedish Research Council) 
through the Oskar Klein Centre. Part of this work was completed at the Niels Bohr Institute, which SV acknowledges for 
hospitality. We are grateful to Miguel Escudero for help regarding the comparison of our model to warm DM. 
SV acknowledges Aurel Schneider, Stuart Wyithe and Francis-Yan Cyr-Racine for useful correspondence, 
and Jes\'{u}s Zavala, Zurab Berezhiani and Miguel Escudero for valuable discussions.

\bibliographystyle{ieeetr}

{\small 
\bibliography{bibliodao}
}

\end{document}